\documentclass[10pt,journal]{IEEEtran}
\usepackage{hyperref}
\usepackage{graphicx}
\usepackage{multirow}
\usepackage{tabularx}
\usepackage{tabu}

\usepackage{amsmath,amssymb}
\usepackage{bm}
\usepackage[]{algorithm2e}

\usepackage[flushleft]{threeparttable}

\usepackage{multirow}

\usepackage{hyperref}
\usepackage{breakurl}
\usepackage{color}

\ifCLASSOPTIONcompsoc
  \usepackage[nocompress]{cite}
\else
  \usepackage{cite}
\fi
\ifCLASSINFOpdf
\else
\fi
\begin{document}
%
\title{Tackling Unit Commitment and Load Dispatch Problems Considering All Constraints with Evolutionary Computation}
%
%
%
%

\author{Michael~Shell,~\IEEEmembership{Member,~IEEE,}
        John~Doe,~\IEEEmembership{Fellow,~OSA,}
        and~Jane~Doe,~\IEEEmembership{Life~Fellow,~IEEE}
\IEEEcompsocitemizethanks{\IEEEcompsocthanksitem M. Shell was with the Department
of Electrical and Computer Engineering, Georgia Institute of Technology, Atlanta,
GA, 30332.\protect\\
E-mail: see http://www.michaelshell.org/contact.html
\IEEEcompsocthanksitem J. Doe and J. Doe are with Anonymous University.}
\thanks{Manuscript received April 19, 2005; revised August 26, 2015.}}

\author{Danilo Vasconcellos Vargas, Junichi Murata and Hirotaka Takano
\thanks{D. V. Vargas and J. Murata are with the Faculty of Information Science and Electrical Engineering, Kyushu University, Fukuoka, Japan (email: vargas@inf.kyushu-u.ac.jp and murata@cig.ees.kyushu-u.ac.jp)}
\thanks{H. Takano is with the Faculty of Engineering, Gifu University, Gifu, Japan (email: takano@gifu-u.ac.jp)}
}

%
%


\markboth{}%
{}
%



\maketitle

\begin{abstract}
Unit commitment and load dispatch problems are important and complex problems in power system operations that have being traditionally solved separately. 
In this paper, both problems are solved together without approximations or simplifications. 
In fact, the problem solved has a massive amount of grid-connected photovoltaic units, four pump-storage hydro plants as energy storage units and ten thermal power plants, each with its own set of operation requirements that need to be satisfied. 
To face such a complex constrained optimization problem an adaptive repair method is proposed. 
By including a given repair method itself as a parameter to be optimized, the proposed adaptive repair method avoid any bias in repair choices.
Moreover, this results in a repair method that adapt to the problem and will improve together with the solution during optimization. 
Experiments are conducted revealing that the proposed method is capable of surpassing exact method solutions on a simplified version of the problem with approximations as well as solve the otherwise intractable complete problem without simplifications. 
Moreover, since the proposed approach can be applied to other problems in general and it may not be obvious how to choose the constraint handling for a certain constraint, a guideline is provided explaining the reasoning behind. 
Thus, this paper open further possibilities to deal with the ever changing types of generation units and other similarly complex operation/schedule optimization problems with many difficult constraints.
\end{abstract}


\IEEEpeerreviewmaketitle

\maketitle





\section{Introduction}

Unit commitment (UC) and load dispatch (LD) are very important problems in power system operations in order to make an appropriate power generation schedule over a specified time period that meets electrical power demand with minimum costs satisfying other relevant constraints \cite{soliman2011modern}. Unit commitment determines which generation units should be run and which should be stopped in each time slot in the period. Load dispatch divides the total amount of power to be generated over individual running units \cite{soliman2011modern}. 
UC and LD are essentially a single problem that finds the optimal power generation plan. However, due to differences in the nature of the problems, solving them as a single problem is difficult. Traditionally, they were solved separately, sequentially or in a hierarchical manner with UC at the higher level and LD at the lower level. Due to this, a UC-LD problem is sometimes simply referred to as a UC problem \cite{momoh2008electric}. 
UC and LD were solved with different solution techniques, respectively. These `separate' solution processes require some extent of approximation or simplification in the problem formulation. This is why efforts are being made to develop better solution techniques although there already exist a number of solution techniques. Especially, much attention are being payed, recently, to evolutionary computation and swarm intelligence techniques \cite{padhy2004unit}, \cite{saravanan2013solution}.
Recently, the nature of the UC and LD problems is changing due to introduction of non-traditional types of generation units, e.g. wind turbines and photovoltaic, and large-scale storage units, which makes it hard for the existing solution techniques to obtain good solutions. The difficulty arises from the complexity of problems, which comes from a larger number of variables and more constraints of various types than the problems used to have.

\begin{figure*}[htbp]
\centering
 \includegraphics[width=\textwidth]{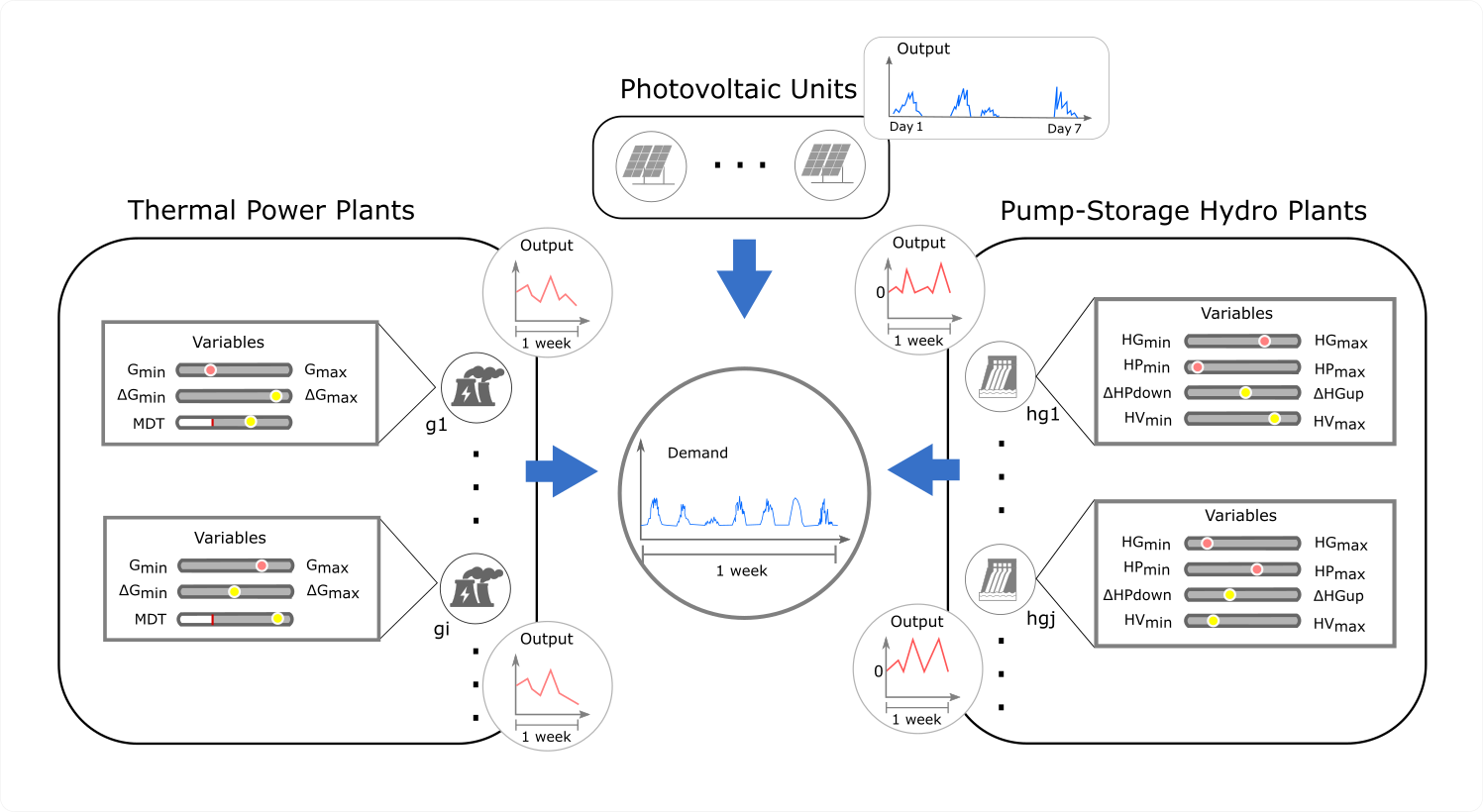}
 \caption{
This paper solves for the first time the combined problem of load dispatch and unit commitment without approximations or simplifications, taking into consideration realistic operation requirements and constraints of a massive amount of grid-connected photovoltaic units, four pump-storage hydro plants as energy storage units and ten thermal power plants.
}
 \label{overview_fig}
\end{figure*}

In this paper a solution technique is proposed which can deal with these more complicated and thus more difficult UC-LC problems without any approximations or simplifications to the standard formulation of the problems (Figure~\ref{overview_fig}).  
The proposed solution technique utilizes the powerful searching capability of differential evolution \cite{storn1997differential} to deal with a large number of variables and treats various types of constraints with a set of newly developed constraint handling techniques, repair-penalty and adaptive repair techniques, combined with existing but effective ones. These constraint handling techniques are not specific to the UC-LD problems but can be applied to various constrained optimization problems.

The proposed solution technique is validated on an example UC-LD problem. 
The target power system includes massive amount of grid-connected photovoltaic and four pump-storage hydro plants as energy storage units in addition to ten thermal power plants. One of the important roles of the pump-storage hydro plants is to store surplus energy generated by the massive photovoltaic when demand is low and to utilize the energy economically. To enhance economical benefit, we have to consider weekly periodicity of the power demand. This forces us to take into account a longer horizon of one week than the one-day horizon that is considered in typical UC and/or LD problems, which results in a huge optimization problem with thousands of variables to be determined. Including pump-storage hydro plants in UC-LD problems also increase complexity of the problems in terms of constraints. A pump-storage hydro plant stores energy by pumping up water to its higher-elevation reservoir. The water level of the reservoir affects how much energy can be generated and how much can be stored by the plant, and it depends on the past operations of the plant. This introduces to the problem equality and inequality constraints with temporal dependence. In this way, the new UC-LD problems are more complicated, and the proposed solution technique will be effective to solve those problems.

\section{Related Work}

The use of optimization techniques in electric power systems has been massively investigated \cite{momoh2008electric}.
Regarding UC optimization problems and their solutions, reviews can be found in \cite{saravanan2013solution}, \cite{padhy2004unit}, \cite{bhardwaj2012unit}, \cite{zheng2015stochastic}. 

Over decades, use of evolutionary algorithms in power systems field has been attracting much attention \cite{soliman2011modern}. 
For UC problems, an early work by Sheblé et al. \cite{sheble1996unit} solved the problem with the genetic algorithm where constraints were handled by penalty methods. 
However, the minimum-down-time constraint, one of the most tricky constraints, was not considered. 
Handling the constraints in UC-LD problems is not easy. 
For example, in \cite{soares2012distributed}, evolutionary algorithms were applied to the problems. 
However, only relatively simple constraints were taken into account and handling the constraints by the repair method may cause some bias on the optimization and give a negative impact on the quality of the search. 
The situation is similar when the LD problems for pumped storage hydro plants considered in \cite{glotic2014optimization}. 
They used the classical differential evolution and an adaptive differential evolution to optimize the unit commitment. There were few constraints taken into account. 
In \cite{zou2016improved}, the valve-point effect is taken into account and the authors used an improved DE for achieving state-of-the-art results. The LD problem considered has, however, few constraints and no hydro plants or photovoltaic units.
Another approach using DE to target a LD problem considering storage hydro plants is presented in \cite{glotic2015short}. In this recent work, an adaptive DE is proposed to solve the problem.
The modeling considered for hydro plants are similar to ours in complexity, however, the thermal plants have only a single constraint associated and there is no photovoltaic units and their associated spinning reserve.
A hybrid approach mixing simulated annealing and ant colony optimization into a new metaheuristic method was applied to the short-term energy resource management problem where intensive use of electric vehicles is present \cite{sousa2014hybrid}. 
In this case, the constraints were all inequalities. To satisfy them, penalties were added to the objective function. 

UC-LD problems for thermal and pumped storage hydro plants were treated in several studies. 
Gollmer et. al \cite{gollmer1999primal} applied the primal and dual method to the problems where no constraints were ignored. Here, however, only linear or piece-wise linear objective functions and constraints were assumed in order to utilize well-established optimization techniques. 
In more recent study \cite{sawa2008daily}, a full set of constraints were considered. However, some of them were relaxed to make use of quadratic programming possible. In summary, handling all the constraints present UC-LD problems is not easy even if evolutionary algorithms are used, and to the best knowledge of the authors, there is no technique that can deal with all the constraints contained in the standard formulation of PV and storage incorporated problems without any approximation or simplification.

\section{Constrained Optimization Problems}

Constrained optimization problems are a type of optimization problems where constraints are present.
Constraints modify the search space making otherwise simple problems extremely complex to solve.
The general formulation of constrained optimization problems is as follows:

\begin{equation}
\begin{split}
	\underset{x}{\operatorname{minimize}}&\quad  f(x),~x \in \mathbb R^n\\
	\operatorname{subject\;to}
	 &\quad g_i(x) \leq 0, \quad i = 1,\dots,m \\
	&\quad h_i(x) = 0, \quad i = 1, \dots,p ,
\end{split}
\end{equation}
where $f(x)$ is the objective function, also called  fitness in evolutionary computation, and $g_i(x)$ and $h_i(x)$ are the two types of constraints. They are called respectively inequalities and equalities. 
$m$ and $p$ define respectively the numbers of inequalities and equalities of the problem.
Thus, for one candidate solution $x$ to be feasible it must satisfy all equalities and inequalities.

Constrained optimization problems are harder than unconstrained ones because the feasible search space itself is not always explicitly given or known.
Constraints can also depend on a number of variables, making almost impossible the prediction of their feasibility.
To deal with these constraint related problems many techniques have been developed, they are discussed in the next Section.

\section{Constraint-Handling Techniques}

Here related constraint-handling techniques used together with evolutionary computation will be described briefly.
An extensive review is out of the scope of this article.

\subsection{Penalty Approaches}
Penalty techniques worsen the fitness of infeasible solutions by adding a penalty function to the fitness.
Penalty functions should be able to make infeasible solutions unlikely to win when competing with feasible solutions.
Therefore, infeasible solutions would coexist with feasible ones but would be discarded in the long run.

There are many types of penalty functions, it is possible to make them depend on the number of generations, i.e., allowing infeasible solutions to violate constraints almost free of penalty in the beginning of evolution while increasing the penalty along the evolution.
It is also possible to modify the penalties using external cooling schemes \cite{joines1994use},
adaptive heuristics \cite{bean1993dual}, another evolutionary algorithm \cite{coello2000use}, parent and children population \cite{oyman1999alternative}, among others.

There are also methods that go to the extreme of discarding any infeasible solutions.
This type of approach is called death penalty \cite{kramer2006three}.

\subsection{Repair Approaches}

Constraint-handling approaches that aim to make infeasible solutions feasible can be divided into two groups \cite{padhye2015feasibility,coello2002theoretical}
\begin{itemize}
	\item Variable-based methods - Methods that test and modify each variable independently to keep the candidate solution inside the feasible region.
	\item Vector-based methods - Methods that when the solution lies in an infeasible region, modify the solution as a vector along a given direction to bring it into feasible space. In this manner, all variables are modified including the ones that were already inside their feasible region.
\end{itemize}

\subsection{Decoder Approaches}

Decoder approaches map a solution from the original search space to another space where solutions are guaranteed to be feasible \cite{eiben2003introduction,coello2002theoretical,koziel1999evolutionary,michalewicz2013solve}.

Some methods define operators and solution representations to force solutions to be feasible \cite{paredis1994co,schoenauer1996evolutionary}.

\subsection{Multi-objective Approaches}

Constraints by themselves can be seen as other objectives to be optimized. These objectives can be easily constructed from constraints by defining a function that measures the constraint violation.
In this view, a constrained single objective problem becomes a multi-objective one and can be handled with multi-objective algorithms such as NSGAII\cite{schoenauer1996evolutionary}, GDE3\cite{kukkonen2005gde3} or SAN\cite{vargas2015general}.
This approach was used by \cite{coello2000treating,jimenez1999evolutionary,parmee1994development,surry1995multi}.



\section{Differential Evolution}
\label{differential-evolution-sec}

Differential Evolution (DE) \cite{storn1997differential} is a global optimization algorithm within the subfield of evolutionary computation.
Similar to other evolutionary algorithms, DE is capable of optimizing functions without knowing the functions to be optimized.
The objective functions themselves are neither required to be continuous nor to be differentiable. 
DE is a fairly simple algorithm based on three procedures (mutation, crossover and selection).
The algorithm is described succinctly in Table~\ref{de_alg} and the explanation of each of its three procedures is done in the subsections below.
\begin{table}[h]
 \centering
\caption{Differential Evolution Algorithm} 
\begin{tabular}{p{8cm}}
\hline
\begin{enumerate}
\item Initialize population with randomly sampled individuals (candidate solutions) 
\item For each generation until the maximum number of generations do:
\begin{enumerate}
\item For each individual in the population do:
\begin{enumerate}
\item Mutation 
\item Crossover
\item Selection
\end{enumerate}
\end{enumerate}
\item Return best candidate solution
\end{enumerate} \\
\hline
\end{tabular}
\label{de_alg}
\end{table}

\subsection{Mutation}

For each candidate solution (vector) in the g-th iteration the mutation is applied by the following equation:
\begin{equation}
v_{i,g+1} = x_{r1,g} + F(x_{r2,g}-x_{r3,g}),
\label{de_mutation_eq}
\end{equation}
where $r1$, $r2$ and $r3$ are indices of randomly selected individuals of the population, which must differ from the individual $i$. $F$ is a parameter which should meet the condition $F \in [0,2]$. 
The generated vector $v_{i,g+1}$ is called mutation vector.

\subsection{Crossover}

During the crossover, a trial vector $u_{i,g+1}$ is created from a combination of the mutation vector $v_{i,g+1}$ and the original vector $x_{i,g}$ is as follows:
\begin{equation}
u_{i,j,g+1} = \left\{\begin{array}{cc}
 x_{i,j,g} & \mbox{if $rand() > CR$ and $j \neq rnd_i$};\\
 v_{i,j,g+1} & \mbox{if $rand() \leq CR$ or $j = rnd_i$},
\end{array} \right. 
\label{de_crossover_eq}
\end{equation}
where $rand() \in [0,1]$ is a uniformly distributed random number, $CR \in [0,1]$ is a parameter, $j$ is the vector component index and $rnd_i$ is a randomly chosen vector index.

\subsection{Selection}

Both $u_{i,g+1}$ and $v_{i,g}$ are evaluated and the vector with better fitness function is kept, forming the next generation vector $x_{i,g+1}$.

\section{Problem Description}
\label{problem_description}


\subsection{Overview}
The target problem is to build an optimal one-week generation schedule in the emerging type of UC-LC problem settings where thermal power plants, photovoltaic (PV) generation units and pumped storage hydro plants as large-scale energy storage exist.

The purpose here is twofold: the first one is to minimize operation costs of the power plants while the second being to avoid wasting the PV unit output power. 
Notice that the operating costs considered here include the fuel costs and the start-up costs of the thermal plants.

When a huge number of PV units are connected to the power system, it can happen that generated output exceeds electricity demand. To avoid wasting this excessive power, the power (energy) must be stored. Here pumped storage hydro plants are used. 
The stored energy must be used to generate power appropriately so that the generation costs of the thermal plants are minimized.

\subsection{Symbols}
The following symbols are used:\\
\begin{tabular}{rp{6cm}}
$t$: & time interval (an hour), (all the suffixes $t$ below indicate values at time $t$),\\
$T$: & set of time intervals $\{0,1,\dots,T_f\}$, \\
$i$: &  index of a thermal plant, (all the suffixes $i$ below indicate values associated to thermal plant $i$), \\
$NG$: & set of indices $i$, \\
$u_{i,t}$: & ($\in \{0,1\}$) thermal plant status, 0: halted, 1: working,\\
$u_{i,t}^{\mathrm{off}}$: & consecutive time length of halted status of a thermal plant up to time $t$,\\
$g_{i,t}$: & thermal plant generating power,\\
$G_i^{\mathrm{max}}$: & maximum output of a thermal plant,\\  
$G_i^{\mathrm{min}}$: & minimum output of a thermal plant,\\  
$\Delta G_{i}^{\mathrm{up}}$: & maximum upward ramp rate of a  thermal plant,\\
$\Delta G_{i}^{\mathrm{down}}$: & maximum downward ramp rate of a thermal plant,\\
$MDT_i$: & minimum down time (minimum required time for a halted plant to restart) of a thermal plant,\\
$SC_i$: & start up cost of a thermal plant,\\
$A_i, B_i$ and $C_i$: & coefficients in fuel cost function of a thermal plant,\\
$j$: &  index of a pumped storage plant, (all the suffixes $j$ below indicate values associated to pumped storage plant $j$)\\
$NHG$: & set of indices $j$, \\
\end{tabular}
\\
\begin{tabular}{rp{6cm}}
$hg_{j,t}$: & pumped storage plant generating power, positive value: generated power, negative value: consumed power for pumping-up,\\
$hv_{j,t}$: & water level of the reservoir for a pumped storage plant,\\
$HG_j^{\mathrm{max}}$: & maximum generating output of a pumped storage plant,\\
$HG_j^{\mathrm{min}}$: & minimum generating output of a pumped storage plant,\\
$HP_j^{\mathrm{max}}$: & ($\ge 0$) maximum consumed power for pumping up,\\
$HP_j^{\mathrm{min}}$: & ($\ge 0$) minimum consumed power for pumping up,\\
$\Delta HG_{j}^{\mathrm{up}}$: & maximum ramp rate of a pumped storage plant in the generator mode,\\
$\Delta HP_{j}^{\mathrm{down}}$: & maximum ramp rate of a pumped storage plant in the pump mode,\\
$HV_j^{\mathrm{max}}$: & maximum water level of the reservoir for a pumped storage plant,\\
$HV_j^{\mathrm{min}}$: & minimum water level of the reservoir for a pump storage plant,\\
$D_t$: & predicted total demand,\\
$pv_t$: & predicted total output from all PV units,\\
$\bm{u}$: & vector that consists of $u_{i,t}, i \in NG, t \in T$,\\
$\bm{g}$: & vector that consists of $g_{i,t}, i \in NG, t \in T$,\\
$\bm{h}$: & vector that consists of $hg_{j,t}, j \in NHG, t \in T$.
\end{tabular}

Some other symbols will also be used. They will be defined where they first appear.

\subsection{Objective function}

The objective function of the problem is defined as follows:
\begin{equation}
\begin{split}
  F(\bm{u},\bm{g},\bm{h})& =\sum_{t \in T}\sum_{i \in NG} [(A_i+B_i \, g_{i,t}
 +C_i \, {g_{i,t}}^2)\,u_{i,t}\\
 &\quad +SC_i\, u_{i,t}\, (1-u_{i,t-1})],
\end{split}
\end{equation}
which is to be minimized. Here, the factor $A_i+B_i \, g_{i,t}+C_i \, {g_{i,t}}^2$ represents the fuel cost of thermal plant $i$, which is a quadratic function of its output, $g_{i,t}$, and is considered only when the plant is working ($u_{i,t}=1)$. 
We include the start-up cost $SC_i$ in the objective function when $u_{i,t}=1$ and $u_{i,t-1}=0$, which means that the plant was halted at the previous time $t-1$ but now at time $t$ is working. 
Note that variable $\bm{h}$ does not appear in the function explicitly. It affects the other variables $\bm{u}$ and $\bm{g}$ through the supply demand balance constraint.

\subsection{Constraints}
\subsubsection{Supply demand balance}
Total power supply and total demand must be equal at any time:
\begin{equation}
 \sum_{i \in NG}g_{i,t}\, u_{i,t}+\sum_{j \in NHG} hg_{j,t}+pv_t=D_t,\:\mathrm{for}\:\forall t.
\end{equation} 
To distinguish clearly our decision variables $u_{i,t}$ and $hg_{j,t}$ from exogenous variables $hv_t$ and $D_t$, the following equivalent expression can be helpful:
\begin{equation}
\label{eq_supply_demand}
 \sum_{i \in NG}g_{i,t}\, u_{i,t}+\sum_{j \in NHG} hg_{j,t}=D_t-pv_t,\:\mathrm{for}\:\forall t.
\end{equation}
Here we term the right hand side, $D_t-pv_t$, as {\em net demand} because it indicates power to be supplied from the thermal and pumped storage plants owned by the power utility company.

When PV units generate huge power in total, the right hand side of equation (\ref{eq_supply_demand}) becomes small. In this situation, it may be necessary to make $hg_{j,t}$ negative (pump mode), which means that the excessive power from the PV units is stored in the reservoirs. 
 
\subsubsection{(Extended) spinning reserve}
\label{spinning_reserve_sec}
We do not know PV unit outputs and demand for sure when we plan the plant operation. Only available are their predicted values, $pv_t, t \in T$ and $D_t, t \in T$, and of course they are prone to errors. To cope with the errors, we introduce the following constraints:
\begin{equation}
\begin{split}
  \sum_{i \in NG}g_{i,t}^{\mathrm{min}}\, u_{i,t}+\sum_{j \in NHG} hg_{j,t}^{\mathrm{min}} 
  \le  (1-\alpha_t)(D_t-pv_t),\\
  \mathrm{for}\: \forall t, \label{lower_reserve}\\
\end{split}
\end{equation}
\begin{equation}
\begin{split}
  \sum_{i \in NG}g_{i,t}^{\mathrm{max}}\, u_{i,t}+\sum_{j \in NHG} hg_{j,t}^{\mathrm{max}}
  \ge  (1+\beta_t)(D_t-pv_t),\\
  \mathrm{for}\: \forall t. \label{upper_reserve}
\end{split}
\end{equation}
Here we introduce the following new symbols:\\
\begin{tabular}{rp{7cm}}
$\alpha_t$: &($\in [0,1]$) the maximum possible decrease in net demand from its predicted value,\\
$\beta_t$: &($\in [0,1]$) the maximum possible increase in net demand from its predicted value,\\
$g_{i,t}^{\mathrm{min}}$: & thermal plant minimum generating output at time $t$,\\
$hg_{j,t}^{\mathrm{min}}$: & pumped storage plant minimum generating output at time $t$,\\
$g_{i,t}^{\mathrm{max}}$: & thermal plant maximum generating output at time $t$,\\
$hg_{j,t}^{\mathrm{max}}$: & pumped storage maximum generating output at time $t$.
\end{tabular}


\subsubsection{Thermal power plants}

Each thermal plant has the maximum and minimum output constraints:
\begin{equation}
 \label{eq_maxmin_thermal}
 G_{i}^{\mathrm{min}}\, u_{i,t} \le g_{i,t}\, u_{i,t} \le G_{i}^{\mathrm{max}},\: \mathrm{for}\: \forall i, \forall t.
\end{equation}
the ramp rate constraints:
\begin{equation}
\label{eq_ramp}
 -\Delta G_i^{\mathrm{down}} \le g_{i,t} - g_{i,t-1} \le \Delta G_i^{\mathrm{up}},\: \mathrm{for}\: \forall i, \forall t.
\end{equation}
and the minimum downtime constraints:
\begin{equation}
\label{eq_mdt}
 \text{If } 0 < u_{i,t}^{\mathrm{off}} < MDT_i \text{ then }\: u_{i,t}=0,\: \mathrm{for}\: \forall i, \forall t.
\end{equation}
which involves the decision variable $u_{i,t}$ over a certain period.

\subsubsection{Pumped storage power plants and their reservoirs}

A pumped storage plant has the maximum and minimum output constraints when working as a generator:
\begin{equation}
\begin{split}
\label{eq_maxmin_pump_gen}
 \text{If }hg_{j,t}>0 \text{ then } 0 \le HG_j^{\mathrm{min}} \le hg_{j,t} \le HG_j^{\mathrm{max}},\\
 \mathrm{for}\: \forall j, \forall t,
\end{split}
\end{equation}
and when working as a pump:
\begin{equation}
\begin{split}
\label{eq_maxmin_pump_store}
 \text{If }hg_{j,t}<0 \text{ then } -HP_j^{\mathrm{max}} \le hg_{j,t} \le -HP_j^{\mathrm{min}} \le 0,\\
 \mathrm{for}\: \forall j, \forall t.
\end{split}
\end{equation}
It also has the ramp rate constraints:
\begin{equation}
\begin{split}
\label{eq_ramp_pump_generation}
 \text{If }hg_{j,t}>0 \text{ then } hg_{j,t}-hg_{j,t-1} \le \Delta HG_j^{\mathrm{up}},\\  \mathrm{for}\:\forall j, \forall t,
\end{split}
\end{equation}
\begin{equation}
\begin{split}
\label{eq_ramp_pump_consumption}
 \text{If }hg_{j,t}<0 \text{ then } hg_{j,t}-hg_{j,t-1} \ge \Delta HP_j^{\mathrm{down}},\\ \mathrm{for}\: \forall j, \forall t.
\end{split}
\end{equation}

Water level of a reservoir changes depending on operation of the pumped storage plant:
\begin{equation}
\label{eq_water_level_changes}
 \varepsilon_j\, hv_{j,t}=\varepsilon_j \, hv_{j,t-1}+\eta_j \, (-hg_{j,t}),\: \mathrm{for}\:\forall j, \forall t.
\end{equation}
The constraints also involve the decision variables $hv_{j,t}$ over a certain period.

Here we use the following new symbols:\\
\begin{tabular}{rp{7.5cm}}
$\varepsilon_t$: &coefficient that converts water level to power (energy),\\
$\eta_j$: &efficiency coefficient of a pumped storage plant.
\end{tabular}{}\\
The reservoir has a limited capacity:
\begin{equation}
\label{eq_water_level}
 HV_j^{\mathrm{min}} \le hv_{j,t} \le HV_j^{\mathrm{max}},\: \mathrm{for}\: \forall j, \forall t.
\end{equation}

Values of some variables at time $t$ depend on the values at time $t-1$. This implies that they also depend on the values at time $t-2$, $t-3$ and so on. Therefore, in theory, the problem must be solved considering an infinitely long time period. This, however, is of course impossible. So we focus on just one week. However, focusing on one week provides us with a solution where all the water stored in the reservoir is used up at the end of the week, which is not a long-term optimal solution. To avoid this, we impose the following constraint on the initial water level, $hv_{j,0}$ and the final water level, $hv_{j,T_f}$:
\begin{equation}
\label{eq_initial_final_water}
 hv_{j,T_f}=hv_{j,0},\:\text{for }\forall j.
\end{equation} 


\section{Solution Representation (i.e., Chromosome Encoding)}

The solution representation used is a concatenation of two matrices, one vector and a positive integer.
The details of the representation are explained below:
\begin{itemize}
	\item Thermal Generator Output Matrix - It is a matrix describing the power generated at a given time by a thermal generator. The value at a given $i,j$-th index of the matrix represents the output of the $i$-th thermal plant at the $j$-th time step. If its value is negative, it means the generator is turned off.
	\item Pump Generator Output Matrix - It is a matrix representing the power generated or consumed by a pumped storage hydro plant at a given time. The value at a given $i,j$-th index of the matrix represents the output of the $i$-th pump generator plant at the $j$-th time step. If its value is negative, it means the generator is consuming power rather than producing it.
	\item Preference Vector ($P$) - This vector defines the order to modify the output of thermal power plants in order to satisfy a given constraint. In this article, this preference vector is used to satisfy the supply-demand constraint.   
	\item Maximum Change per Step ($MaxC$) - While the chromosome is being repaired this variable controls the maximum output change of one plant in one step.  
\end{itemize}

Note that the values determined by the chromosome are {\em pre-repair} values that might be repaired later to fit constraints. 
The details of how a chromosome is repaired will be explained later on.

\begin{figure*}[!ht]
\centering
 \includegraphics[width=0.9\textwidth]{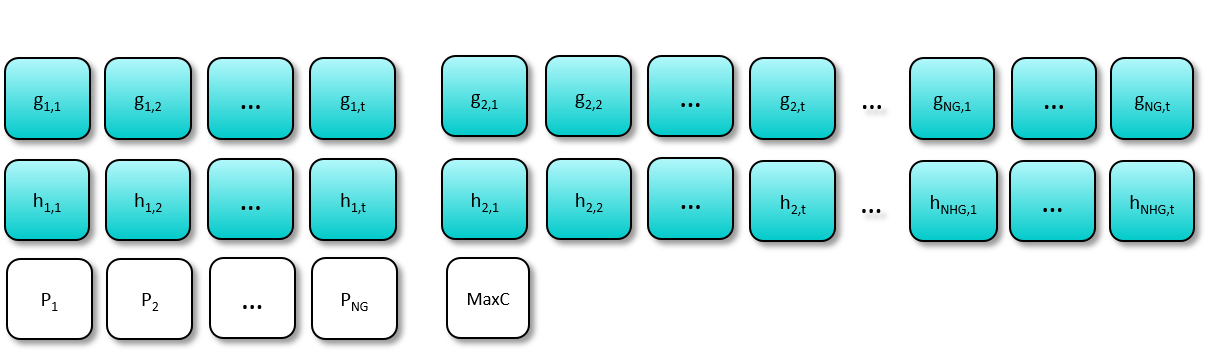}
 \caption{Solution representation. Colored blocks are problem parameters while white blocks are repair related parameters to be optimized by the algorithm. $P$ is a preference vector and $MaxC$ is the maximum change per step for the repair procedure.}
 \label{encoding}
\end{figure*}

\section{Proposed Method: Overview}
\label{overview}

In this paper, a global optimization algorithm is combined with several new and existing constraint handling techniques.
DE algorithm is chosen as the global optimization algorithm \cite{storn1997differential} for its simplicity and good performance.
The UC-LD problem is a complex optimization problem.
Therefore, it is reasonable to choose DE, which is capable of searching in its huge search space.

In addition to the DE algorithm, appropriate constraint handling mechanisms should be employed.
Adding penalty functions is perhaps the simplest form of constraint handling mechanism which is enough to deal with some form of inequalities.
However, penalty functions are not effective to deal with equalities because in general there are much more infeasible solutions.
Nevertheless, given a candidate solution, satisfying equalities and single-variable inequalities often requires only a trivial repair mechanism.
This is why we use a series of repair mechanisms to rebuild a candidate solution and consequently satisfy many equalities and single-variable inequalities (Section~\ref{repair}).
The hardest part of designing repair mechanisms for most of the equalities comes from deciding the best order as well as repair mechanisms that are as mutually independent as possible to avoid violating previously satisfied constraints (i.e., avoid undoing the work of previous repair mechanisms).

Having said that, there are some exceptions though.
For some equalities of the problem described (Section~\ref{problem_description}), satisfying them is either (a) difficult or (b) prone to induce biases by repairing mechanisms (the repaired solution satisfies the constraints but is not necessarily the best or a recommended one).
To solve the cases (a) and (b) we adopt respectively a repair-penalty mechanism (i.e., a repair mechanism with penalty function) and a adaptive repair mechanism.
These two approaches are briefly described below:
\begin{itemize}
	\item Repair-penalty mechanism -  An additional penalty function is included to repair mechanisms when constraints are left unsatisfied. 
		This penalty function is set to zero if the repair mechanism can satisfy the constraint, only penalizing a candidate solution in which the mechanism cannot repair.
	\item Adaptive repair mechanism - Instead of setting a fixed and biased repairing, an unbiased procedure is developed.
\end{itemize}

In the next section, the details about the repair and evaluation procedures are given.

\section{Proposed Method: Repair and Evaluation}
\label{repair}



Before evaluating a candidate solution, it is repaired by a series of methods that try to satisfy as much as possible the constraints and/or penalized when necessary.
Notice that the order in which constraints are satisfied is important as changes in this order may alter the result.
In fact, modifying a candidate solution to satisfy one constraint may turn this candidate solution infeasible in previously satisfied constraints.
Therefore, careful attention should be paid to how to repair a candidate solution as well as to the order in which these repairs occur.
With the above in mind, below we describe the procedures used to repair single variable inequality constraints and in which order they are executed:
\begin{enumerate}
\item Satisfy ramp rate constraints for thermal plants (Equation~\ref{eq_ramp}) - Anything exceeding the maximum ramp rate is set to its maximum value;
\item Satisfy maximum/minimum value for thermal plants (Equation~\ref{eq_maxmin_thermal}) - Anything exceeding the minimum or maximum is set to its respective minimum/maximum value.
\item Satisfy minimum downtime constraints of thermal plants (Equation~\ref{eq_mdt}) - If a given plant violates the constraint, it is turned off for as many time steps as necessary to satisfy the constraint.
\item Satisfy maximum/minimum generation output for pumped storage plant (Equation~\ref{eq_maxmin_pump_gen}) - Anything exceeding the maximum/minimum is set to its maximum/minimum value for pumped storage plants when working as a generator (i.e., positive value). 
\item Satisfy maximum/minimum power consumption for pumped storage plant (Equation~\ref{eq_maxmin_pump_store}) - Same as above but for pumped storage plants when working as a pump (i.e., negative value).
\item Satisfy ramp rate limits for power storage hydro plant's generation (Equation~\ref{eq_ramp_pump_generation}) - Any increase in generation that exceeds the ramp rate is set to its maximum ramp rate.
\item Satisfy ramp rate limits for power storage hydro plant's consumption (Equation~\ref{eq_ramp_pump_consumption}) - The same as above, but for the increase in consumption rather than generation.
\item Satisfy maximum/minimum water levels (Equation~\ref{eq_water_level}) - If a given power storage plant's consumption or generation would cause its water level to go above or below its maximum/minimum water level according to Equation~\ref{eq_water_level_changes}, the consumption or generation power is set to satisfy the maximum/minimum water level. 
\end{enumerate}


By repairing the chromosome with the mentioned procedures, the candidate solution can now easily satisfy many constraints without inserting any kind of bias. This happens because the satisfied constraints are simple in nature, i.e., there is only one procedure for satisfying them.

However, there are still some constraints which need to be satisfied.
Among the remaining constraints there are the supply demand balance constraint (Equation~\ref{eq_supply_demand}), the constraint on the initial and final water levels (Equation~\ref{eq_initial_final_water}) and the spinning reserve constraint (Equation~\ref{lower_reserve} and~\ref{upper_reserve}), i.e., two equalities and one inequality.

\subsection{Mechanism to Repair the Supply Demand Balance Constraint}
\label{adaptive_repair_mechanism}

To satisfy the supply demand balance constraint a newly developed repair method is used, increasing/decreasing the output value of power plants in sequence until the supply matches demand.
In this repair mechanism the order of repairs for each power plant matters.
Therefore, fixing the order gives a strong and undesirable bias.
Moreover, another bias would be inserted if the repair's maximum change per step $MaxC$ is fixed.

To avoid these biases a new type of constraint handling mechanism called Adaptive Repair Mechanism is proposed.
This new constraint handling mechanism, instead of fixing the order of repairs and $MaxC$, allows the repair mechanism to be controlled by both a vector of parameters called preference vector and $MaxC$.
The values inside the preference vector set the priority of repair for each thermal plant, i.e., thermal plants with higher values in the preference vector are repaired first.
The preference vector and $MaxC$ are added to the parameters to be optimized.
Consequently, every candidate solution has its own way of repairing itself.
In fact, an interesting byproduct of this approach is that the way of repairing itself is optimized as well.

Aside from the repair of values to set supply equal to demand, it is also necessary to verify if ramp constraints continue to be satisfied afterwards.
In case they are not satisfied, a simple repair mechanism described before is employed again to satisfy them.
Actually, it is possible that even after the repair mechanism supply does not equal demand.
If this happens, a huge penalty is added to the evaluation of the candidate solution.
The value of this huge penalty is equal to the difference between an ideal value that could satisfy the constraint and the current value multiplied by the $supplyDemandCofficient$.
The detailed description of the repair mechanism can be seen in Algorithm~\ref{supply_demand_alg}.
The $supplyDemandCoefficient$ is a value that starts as zero and is constantly increasing during optimization.
Its maximum value is given by $MaxSupplyDemandCoefficient$.

\begin{algorithm}[h]
\caption{Supply Demand Balance Constraint Repair Algorithm} 
\For{every time step and while $diff \ne 0$}
{
	$diff$ = supply demand difference\;
	
	\For{$MaxAdjustments$ iterations and while $diff \ne 0$}
	{
		\For{all thermal generators}
		{
			Modify thermal generator\;
			Update $diff$ to the current supply demand difference\;
		}
		(\tcc{Notice that all thermal generators are passed in the order given by the preference vector})
	
	}
	
	\If{$diff \ne 0$}
	{
		Add a huge penalty\;
	}

	Repair thermal ramp rate constraints (Equation~\ref{eq_ramp})
}
\label{supply_demand_alg}
\end{algorithm}

\subsection{Mechanism to Repair the Initial and Final Water Level Constraint}
\label{initial_final_water}

The constraint on the initial and final water levels (Equation~\ref{eq_initial_final_water}) is also an equality which makes it more suitable to be tackled with repair mechanisms.
In fact, there is an easy repair mechanism which consists of starting from the last time step $t$ and going backwards in time updating the output of pumped storage hydro plants.
In every update, the output of pumped storage hydro plants are modified to set the difference between initial and final water levels closer to zero.
By going back in time, we guarantee that the modification will not make it surpass, in subsequent time steps, the maximum or minimum water levels. 

In case the repair mechanism is not able to correct the candidate solution, a penalty ($WaterLevelPenalty$) is added to the objective.
The penalty is defined as:
\begin{equation}
\begin{split}
\label{eq_water_penalty}
WLP_j =  |hv_{j,T_f} - hv_{j,0} | ,\:\text{for }\forall j, \\
WaterLevelPenalty =  waterCoefficient\sum_{j \in NHG} WLP_j,
\end{split}
\end{equation} 

The $waterCoefficient$ is similar to $supplyDemandCoefficient$ in that it starts as zero and is constantly increasing during optimization to a maximum value of $MaxWaterCoefficient$.
For each generation of the optimization algorithm, the $waterCoefficient$ and $supplyDemandCoefficient$ increases their value by $coefficientStep$ parameter.

\subsection{Spinning Reserve's Penalty Function}

The spinning reserve constraint (Equations~\ref{lower_reserve} and~\ref{upper_reserve}) is included in the problem in the form of a penalty function.
The penalty for the spinning reserve constraint ($SpinningReservePenalty$) is defined as:
\begin{equation}
\begin{split}
  S1_t = \sum_{i \in NG}g_{i,t}^{\mathrm{min}}\, u_{i,t}+\sum_{j \in NHG} hg_{j,t}^{\mathrm{min}} 
  -  (1-\alpha_t)(D_t-pv_t),\\
  \mathrm{for}\: \forall t, \\
  S2_t = (1+\beta_t)(D_t-pv_t) - 
  \sum_{i \in NG}g_{i,t}^{\mathrm{max}}\, u_{i,t}+\sum_{j \in NHG} hg_{j,t}^{\mathrm{max}} \\
  \mathrm{for}\: \forall t.\\
	SpinningReservePenalty_t = \max\{S1_t, 0\} + \max\{S2_t,0\} \\
\end{split}
\end{equation}

\subsection{Bridging gap between pre-repair solutions and post-repair solutions}

Recall that the chromosomes (candidate solution) are only the pre-repair values.
They are, when necessary, repaired and/or penalized and the fitness is evaluated for the post-repair values. 
To bridge the gap between the pre-repair solutions expressed by chromosomes and their post-repair solutions, every one in $RC$ function evaluations a chromosome is set to the repaired values ($RC$ is set to $10000$ in all experiments).
Notice that a low value of $RC$ would make infeasible solutions disappear rapidly from the population.
This can be dangerous because the space of feasible solutions are much smaller and disconnected, therefore premature convergence is very likely to occur.

\subsection{A Guideline for Building Repair Mechanisms Independent of the Problem}
\label{guidelines}

The optimization method together with its repair mechanisms can be applied to any other optimization problem.
This section aims to aid the construction of repair mechanisms in general.
\begin{itemize}
\item Separate constraints into multivariable ones and single-variable ones.
\item (a) For single-variable constraints do:
	\begin{enumerate}
	\item Check - If the constraint's variable depend on other unsatisfied constraints, treat this as a multivariable constraint and go to item (b) or (c) if the relationship between the constraints is respectively an equality of inequality.
	If the constraint's variable does not depend on other unsatisfied constraints go to the next step;
	\item Solve - Use a greedy procedure to satisfy it (e.g., if a value exceeds the maximum set it to the maximum);
	\item Loop - Go to the next unsatisfied single-variable constraint and repeat the first step 'Check'.
	\end{enumerate}
\item (b) For multivariable inequalities do:
	\begin{enumerate}
	\item Add a Penalty - Multivariable inequalities accept many solutions, i.e., probably it is not difficult to go from feasible to infeasible and vice-versa. 
	\end{enumerate}
\item (c) For multivariable equalities do:
	\begin{enumerate}
	\item Special Solution - Search for a special solution that could solve the multivariable equality constraint without adding bias. For example, solving the multivariable equality constraint backward in time (see Section~\ref{initial_final_water} for an example). If such unbiased solution is not found go to the next step.
	\item Add Adaptive Repair Mechanism - Add a adaptive repair mechanism to avoid bias and at the same time be certain that the equality will be satisfied. The general idea of an adaptive repair mechanism is described in Section~\ref{overview} and an example is given in Section~\ref{adaptive_repair_mechanism}.
	\end{enumerate}
\end{itemize}

\section{Experiment 1: Comparison with Another Approach In A Simplified UC-LD Problem}

\subsection{Problem Settings (Simplified Version)}
\label{problem_settings}

This section explains the problem settings used in the first experiments in which the proposed algorithm is compared with an exact approach. 
Since the exact approach can not face the electrical dispatch (UC-LD) problem considering all constraints, a simplified version of the problem described in Section~\ref{problem_description} is considered.

In the simplified version of the problem, it is assumed that the parameter $\Delta t$ and the variables $\Delta G^{up}_i$ and $\Delta G^{down}_i$ satisfy the conditions shown below.
This assumption eases the difficulty that comes from the large-scale mixed-integer nature of the problem.
\begin{eqnarray*}
& \Delta t \ge MDT_i  & \forall i \\
& \Delta G^{up}_i \ge G^{max}_i - G^{min}_i & \forall i  \\
& \Delta G^{down}_i \ge G^{max}_i - G^{min}_i  & \forall i 
\end{eqnarray*}


Under these assumptions, the target problem can be partitioned into the following three stages and solved by an enumeration method followed by a quadratic programming (QP) solver.
\begin{enumerate}
\item First, all the feasible UC solution candidates are enumerated and their optimal output shares are calculated assuming no pumped storage hydro plants.
If the sum of electricity generated by the thermal and PV units exceeds the power consumption, we can identify how much power becomes surplus at which time in this stage.
\item Afterwards, the QP solver derives the optimal pumped storage hydro operation to minimize fuel costs of the thermal units under the condition that the surplus power must be canceled.
In this stage, the UC solution has been fixed, and we assume that the operating thermal units are approximated as one large-scale thermal unit ($g_t = \sum_{i \in NG} g_{i,t}u_{i,t}$).
The vector $\bm{h}$ influences the fuel cost of thermal units through the relationship $\Delta g_t = -h_t$.
\item Finally, the output shares of thermal units are recalculated in consideration of the determined schedule of $\bm{u}$ and $\bm{h}$.
This procedure does not guarantee the global optimality of solutions but can provide us with stable and consistent solutions.
\end{enumerate}

\begin{figure}[!ht]
\centering
 \includegraphics[width=1.0\columnwidth]{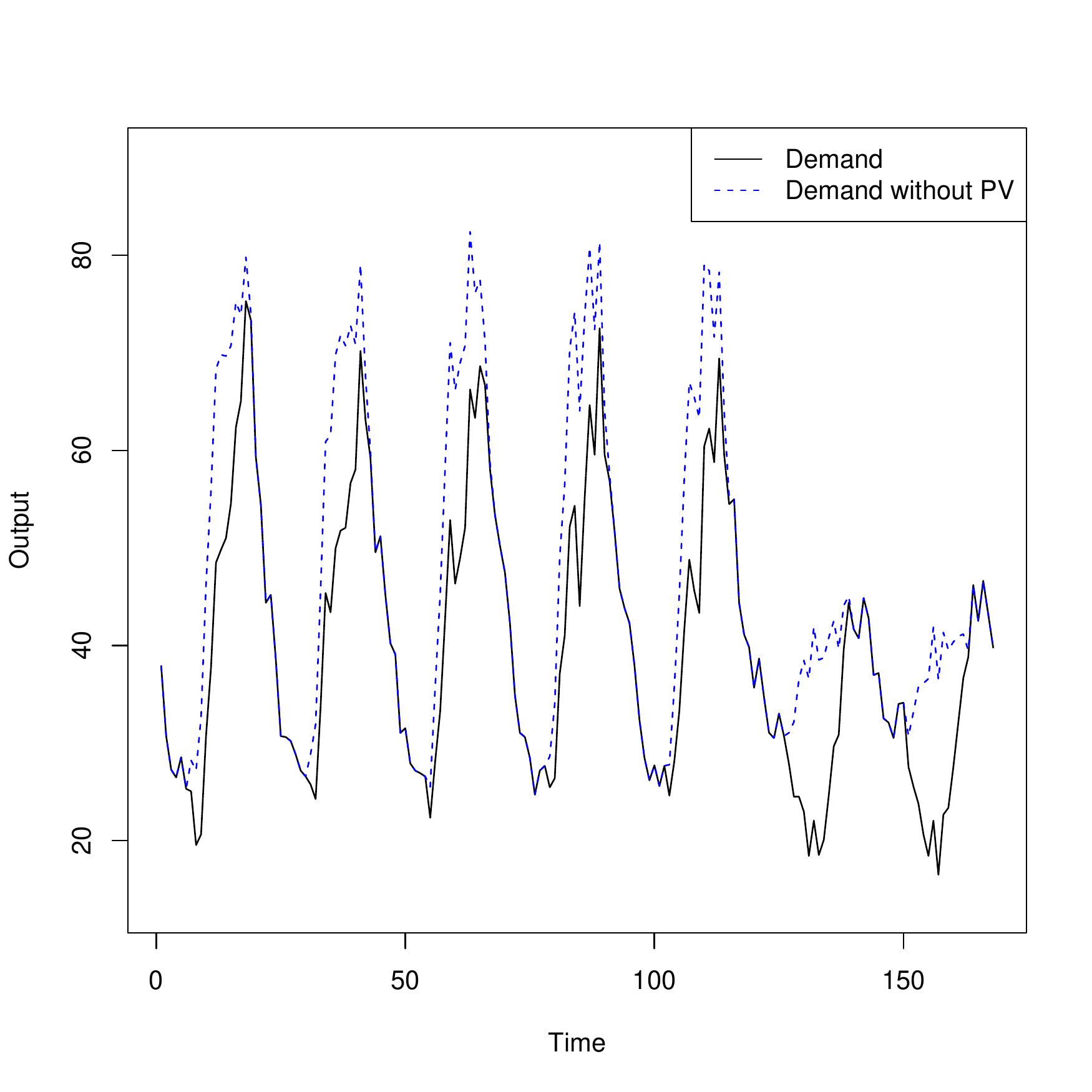}
	\caption{Plot of the real demand (demand minus energy generated by photovoltaic units) and without photovoltaic units.}
 \label{problem_plot}
\end{figure}

The numerical simulation model has $10$ thermal generators each with their own set of parameters and one aggregated PV unit. 
The predicted values of demand and PV output are given, and their difference has to be compensated by the thermal and pump-storage hydro plants (Figure~\ref{problem_plot}).
Naturally, the predicted values include uncertainty and that is why the spinning reserve constraint is taken into account (Equations~\ref{lower_reserve} and~\ref{upper_reserve}).
Tables~\ref{thermal_parameters} and~\ref{thermal_cost_parameters} describe respectively the general settings and the cost related settings of the thermal generators.
To enable both methods to be compared only in terms of thermal plants' output found and their respective cost, both methods are applied to Stage 3 assuming the results have already been obtained in Stages 1 and 2.




\begin{table*}
\centering
\caption{Thermal Generators' General Settings}
\begin{tabu} to \textwidth { |X|X[c]|X[c]|X[c]|X[c]| }
	\hline
	Generator Number & Minimum Uptime (hour) & Minimum Downtime (hour) & Maximum Output & Minimum Output  \\ \hline
	Generator 1 & 8 & 10 & 11 & 11  \\ \hline
	Generator 2 & 6 & 8 & 11 & 11  \\ \hline
	Generator 3 & 7 & 10 & 7 & 1  \\ \hline
	Generator 4 & 8 & 8 & 11 & 3.3  \\ \hline
	Generator 5 & 9 & 5 & 5.8 & 1  \\ \hline
	Generator 6 & 8 & 8 & 11 & 3.3  \\ \hline
	Generator 7 & 9 & 5 & 5.8 & 1  \\ \hline
	Generator 8 & 8 & 8 & 7 & 1  \\ \hline
	Generator 9 & 8 & 8 & 7 & 1  \\ \hline
	Generator 10 & 6 & 7 & 5 & 1  \\ \hline
\end{tabu}
\label{thermal_parameters}
\end{table*}

\begin{table*}[!ht]
\centering
\caption{Thermal Generators' Cost Related Parameters}
\begin{tabu} to \textwidth { |X|X[c]|X[c]|X[c]|X[c]| }
	\hline
	Generator Number & Startup Cost & Cost Coefficient A & Cost Coefficient B & Cost Coefficient C \\ \hline
	Generator 1 & 1 & 0.01 & 0.5 & 0.01 \\ \hline
	Generator 2 & 3 & 0.01 & 0.5 & 0.01 \\ \hline
	Generator 3 & 0.8 & 1.17 & 2.4 & 0.04 \\ \hline
	Generator 4 & 8 & 6.05 & 1.8 & 0.063 \\ \hline
	Generator 5 & 1 & 0.01 & 4.2 & 3 \\ \hline
	Generator 6 & 8 & 6.05 & 1.9 & 0.063 \\ \hline
	Generator 7 & 8 & 3 & 3.9 & 0.01 \\ \hline
	Generator 8 & 2 & 2.1 & 5 & 0.038 \\ \hline
	Generator 9 & 2 & 2.1 & 5 & 0.038 \\ \hline
	Generator 10 & 0.3 & 2 & 5 & 0.05 \\ \hline
\end{tabu}
\label{thermal_cost_parameters}
\end{table*}

\begin{table*}[!ht]
\centering
\caption{Hydropower Plants' Settings (Only Considered in Full Version of the Problem)}
\begin{tabu} { |X[l]|X[c]|X[c]|X[c]|X[c]|X[c]| }
	\hline
	Plant Number & Maximum Output & Maximum Consumption & System Efficiency & Maximum Water Reservoir & Conversion Coefficient \\ \hline 
	Generator 1 & 2.5 & 2.5 & 80 & 100 & 10 \\ \hline
	Generator 2 & 2.5 & 2.5 & 80 & 100 & 10 \\ \hline
	Generator 3 & 2.5 & 2.5 & 80 & 100 & 10 \\ \hline
	Generator 4 & 2.5 & 2.5 & 80 & 100 & 10 \\ \hline
\end{tabu}
\label{hydro_parameters}
\end{table*}

\subsection{Comparison}

\begin{figure}[!ht]
\centering
 \includegraphics[width=1.0\columnwidth]{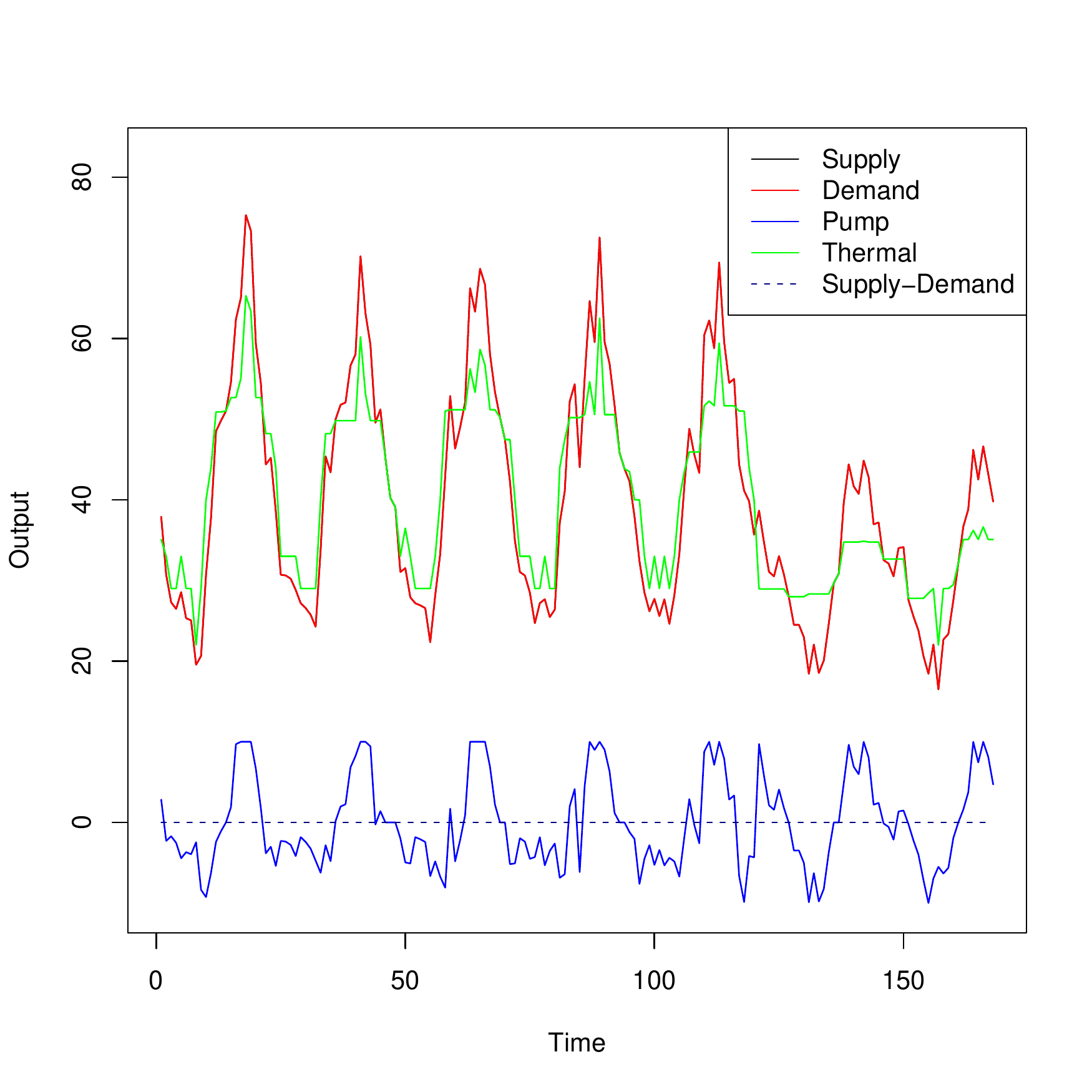}
 \caption{Supply demand curve for the proposed method (with cost $11761$) and the exact method (with cost $12049$) on the simplified problem. Notice that since the pump-storage hydro plant operation has been fixed, the supply demand curve must be the same for both methods if all constraints are satisfied.}
 \label{takano_plot}
\end{figure}

The objective of this section is to compare the algorithm with an exact approach.
The final exact solution to the simplified problem gives the minimum cost of $12049$ (Figure~\ref{takano_plot}). 


Regarding the proposed method, it achieves a cost of $11761.1$ surpassing the exact result.
This means that the solution found satisfy all constraints and is $2.3\%$ more efficient.
To find this result the method was run $30$ times.
Notice that a better solution than the exact approach is possible because there are a series of approximations (e.g., using one single large-scale thermal unit) which are employed by the exact solution in order to solve the problem.
In fact, every run of the proposed method achieves on average $11884$ which is still better than the exact result.
Table~\ref{exp1_table} shows the complete statistics.
The proposed method uses the same parameters for both this simplified experiment and the experiment containing all constraints explained in the next section. Parameters are shown in Table~\ref{parameters2}.

\begin{figure}[!ht]
\centering
 \includegraphics[width=0.7\columnwidth]{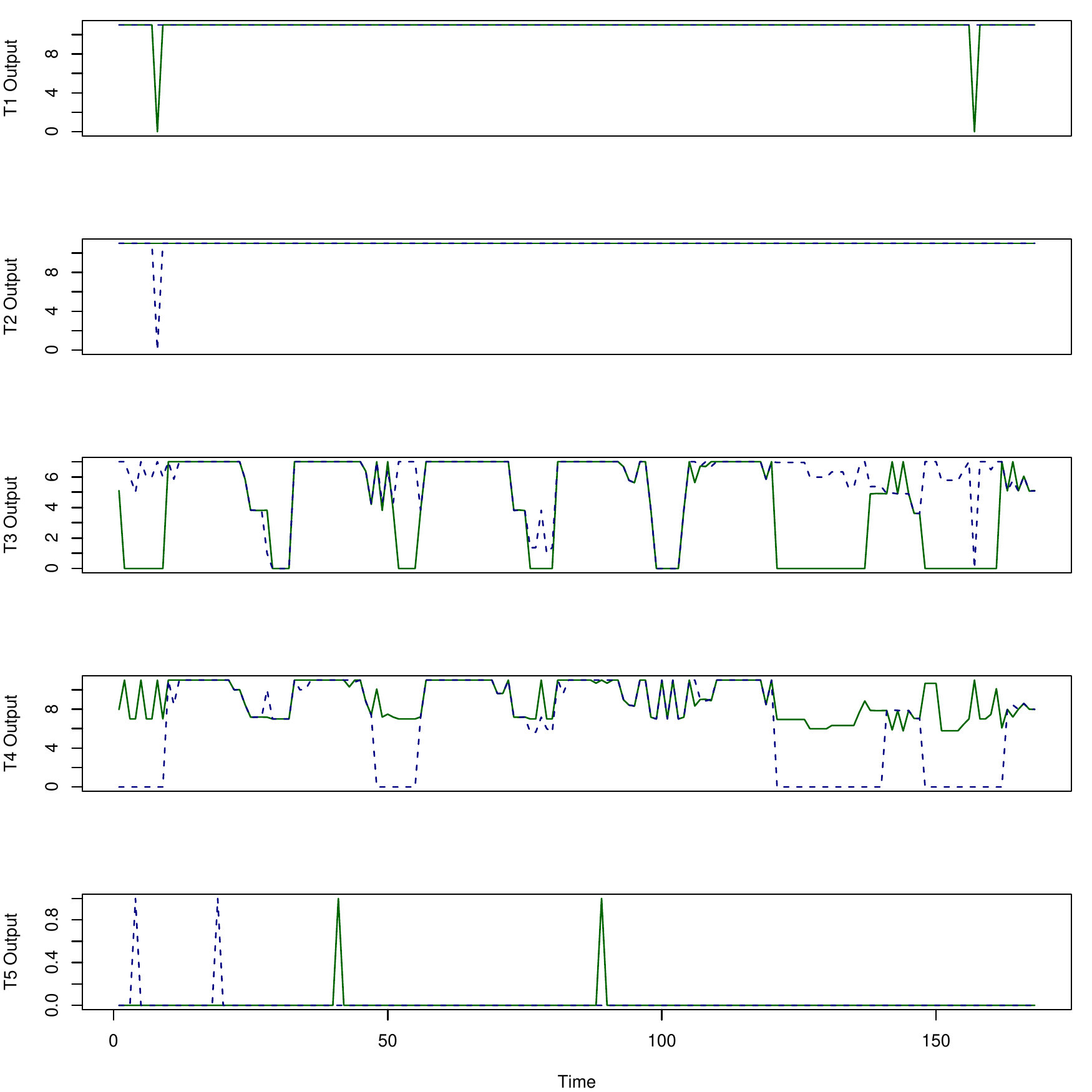}
 \includegraphics[width=0.7\columnwidth]{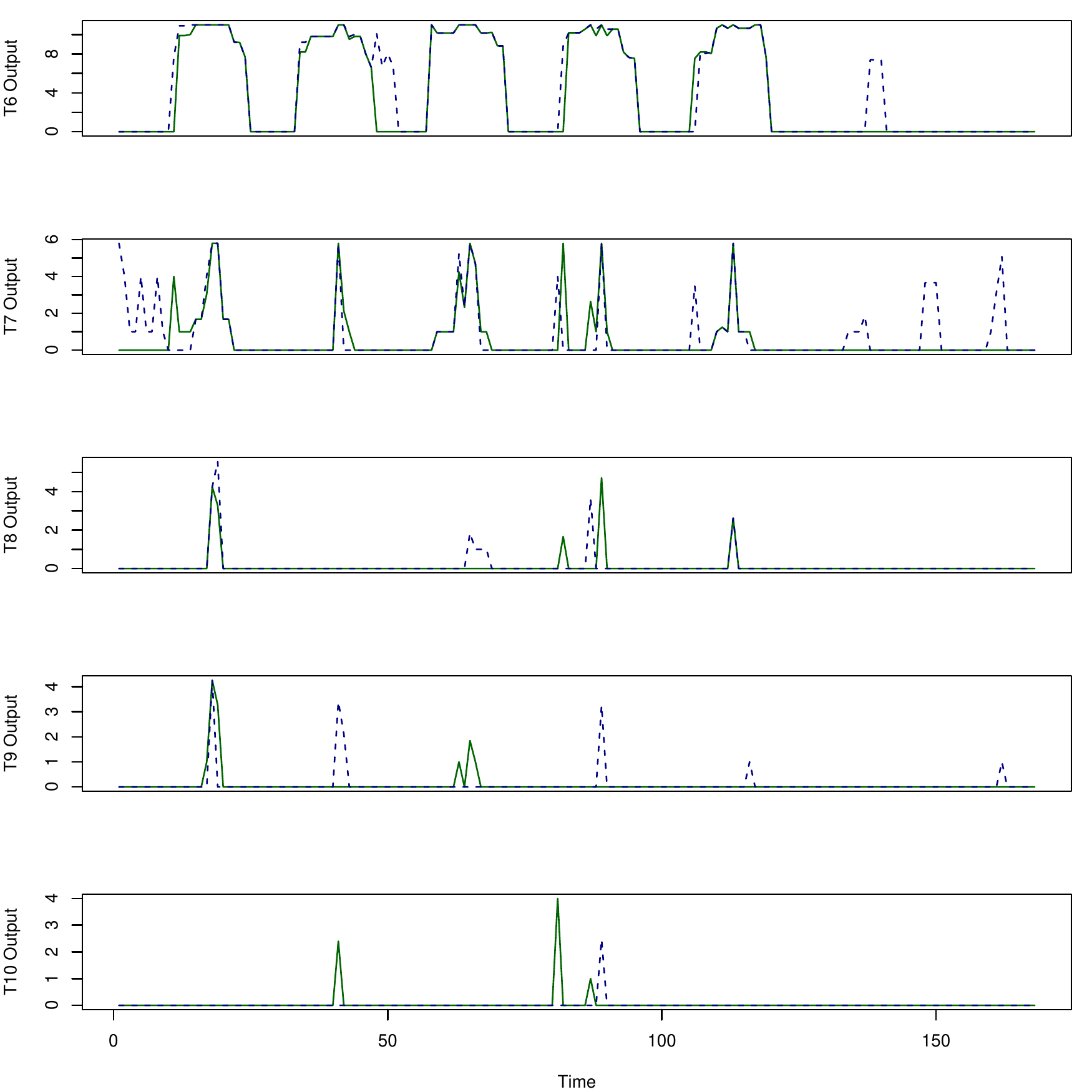}
 \caption{Thermal output comparison between the proposed method (green solid line) and the analytical approach (blue dashed line).}
 \label{takano_my_version}
\end{figure}

Figure~\ref{takano_my_version} shows the thermal plants' outputs for both methods. 
Although both methods are very distinct, they share many similarities in their solutions. 
For example:
\begin{itemize}
	\item Thermal plants $1$ and $2$ are set to their maximum output for almost the whole period.
	\item Thermal plants $5,7,8,9$ and $10$ are almost not used. They exhibit some pulse patterns.
	\item Thermal plants $3,4$ and $6$ are switched on for long periods with some intervals.
\end{itemize}

\begin{table}[!ht]
\centering
\caption{Results of the proposed method in the simplified Experiment~1. Results are averaged over $30$ runs.}
  \begin{threeparttable}
\begin{tabular}{ |l|l| }
	\hline
 	Variable Name & Value \\ \hline
	Best Solution & $11761$  \\
	Solutions With Penalty $< 0.01$ & $13\%$  \\
	\hline
	\hline
 	Variable Name & Mean (Standard Deviation)* \\ \hline
	Cost &$ 11927.4 (203.5)$ \\
	Fitness &$ -11927.4 (203.5)$ \\
	Total Penalty &$ 0.00( 0.00)$ \\
	Spinning Reserve Penalty & $0.00 (0.00)$ \\
	\hline
\end{tabular}
     \begin{tablenotes}
      \small
      \item Fitness and cost have the same absolute value because the penalty is zero. Mean and standard deviation of the final solutions which satisfied all constraints ($13\%$ of the runs).
    \end{tablenotes}

\end{threeparttable}
\label{exp1_table}
\end{table}

\section{Experiment 2: Electrical Dispatch Problem Considering All Constraints}

In this section the same problem is tackled considering all constraints and the hydro power plants' output is to be determined (not fixed).
The following sections describe the parameters settings and discuss the achieved results.

\subsection{Problem Settings (Full Version)}

The problem settings are the same as the one mentioned in Section~\ref{problem_settings}.
There are 4 hydro power plants.
The parameters of all hydro power plants are shown on Table~\ref{hydro_parameters}. 
Regarding the algorithm, its parameters are described in Table~\ref{parameters2}.

\begin{table*}[!ht]
\centering
\caption{Parameters} 
  \begin{threeparttable}
\begin{tabular}{ |l|l| }
	\hline
	Parameter & Value \\ \hline
	  Population size & $2000$ \\
	  Maximum generations & $80000$ \\
	  Maximum adjustment iterations ($MaxAdjustments$) & $10$ \\	
	  $MaxSupplyDemandCoefficient$ & $1000$ \\
	  $MaxWaterCoefficient$ & $100$ \\
	  $coefficientStep$ & $0.025$ \\
	\hline
	\hline
	 Differential Evolution Parameters & Value \\ \hline
	 CR & $0.8$ \\ 
	 F & $random(0,1)*$ \\
	 Initial values & $random(-10,10)*$ \\
	\hline
\end{tabular}
     
     \begin{tablenotes}
      \small
      \item Random($a$,$b$) means that a value is chosen by sampling from a uniform distribution with maximum value $a$ and minimum value $b$.
    \end{tablenotes}
  \end{threeparttable}

\label{parameters2}
\end{table*}

\subsection{Results}

\begin{figure}[!ht]
\centering
 \includegraphics[width=1.0\columnwidth]{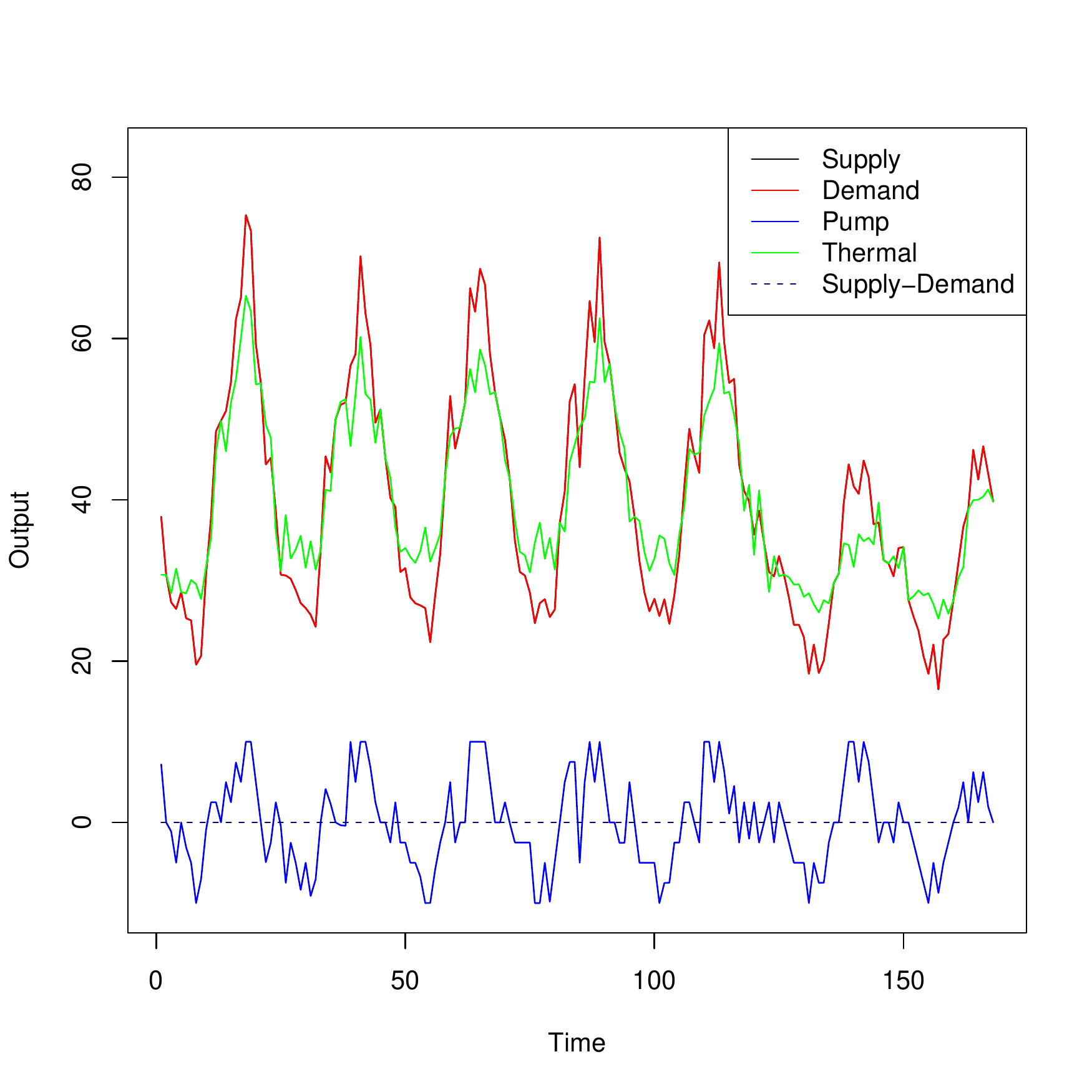}
 \caption{Supply demand curve of the best solution found which has a thermal cost of $12650$ while satisfying all constraints.}
 \label{supply_demand}
\end{figure}

The proposed method is able to solve for the first time the complex UC-LD problem considering all constraints.
The best solution found satisfies all constraints while having a thermal cost of $12650$.
The results of running the proposed algorithm on the problem in question are shown on Table~\ref{results}.
Figure~\ref{supply_demand} shows the supply demand curve of the best result out of $30$ runs.

This is made possible by using the previously described constraint handling techniques which aid the search by correcting candidate solutions.
Consequently, candidate solutions are mapped to a feasible region of the search space, improving searching time and avoiding many local minima.
The new constraint-handling technique proposed is specially important because it allows the search itself to define how corrections are to be made.

\begin{table}[!ht]
\centering
\caption{Results of the proposed method in Experiment~2. Results are averaged over $30$ runs.}
\begin{threeparttable}
\begin{tabular}{ |l|l| }
	\hline
 	Variable Name & Value \\ \hline
	Best Solution & $12650$  \\
	Solutions With Penalty $< 0.01$ & $36\%$  \\
	\hline
	\hline
 	Variable Name & Mean (Standard Deviation)* \\ \hline
	Cost &$ 14360.1 ( 1283.6 )$ \\
	Fitness &$ -14360.1 ( 1283.6)$ \\
	Total Penalty & $0.00 (0.00)$ \\
	Spinning Reserve Penalty & $0.00 (0.00)$ \\
	\hline
\end{tabular}
     \begin{tablenotes}
      \small
      \item Fitness and cost have the same absolute value because the penalty is zero. Mean and standard deviation of the final solutions which satisfied all constraints ($36\%$ of the runs).
    \end{tablenotes}
\end{threeparttable}
\label{results}
\end{table}

\begin{figure}[!ht]
\centering
 \includegraphics[width=1.0\columnwidth]{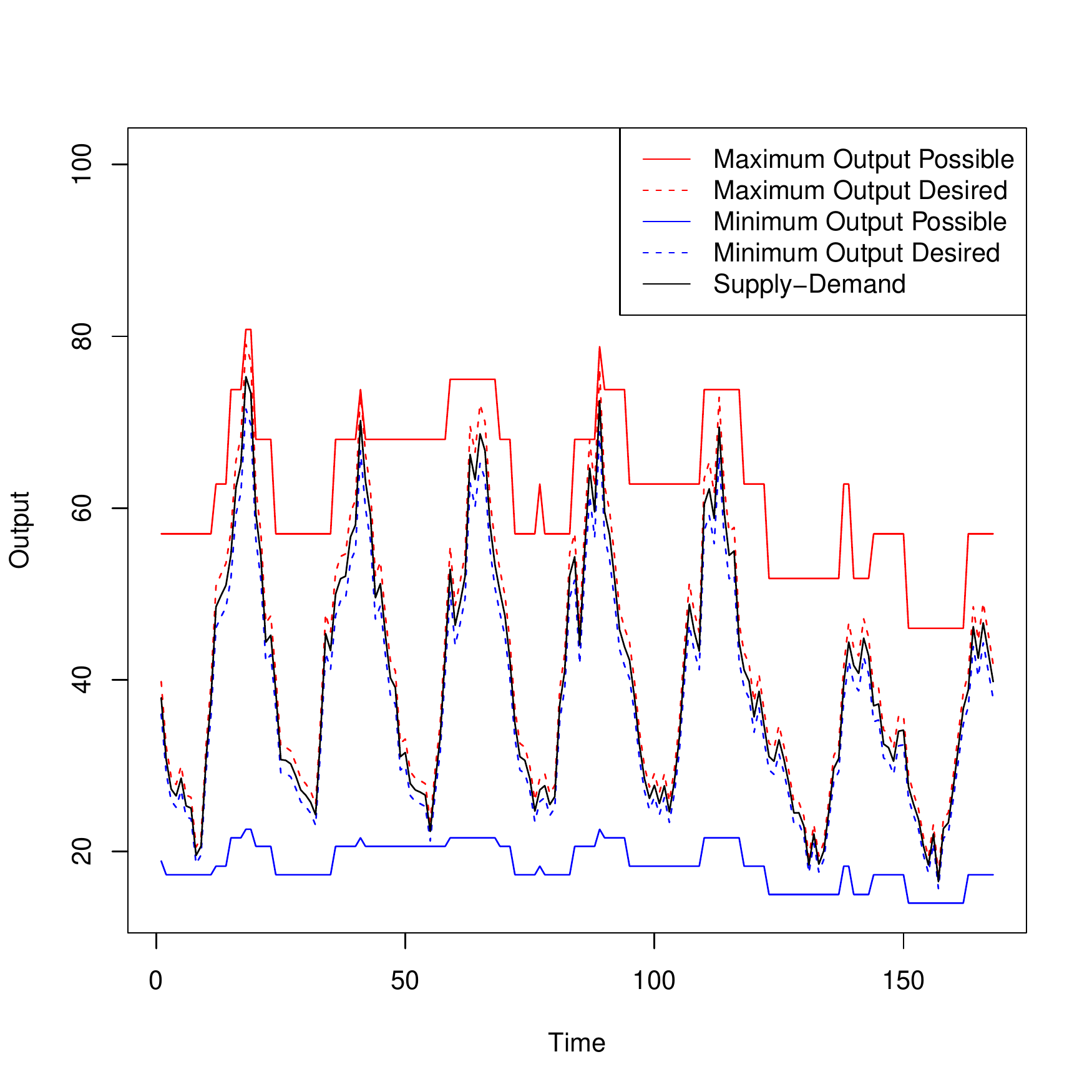}
 \caption{Plot of the maximum and minimum output that the best solution is able to produce together with the desired maximum/minimum output defined by the spinning reserve constraint.}
 \label{spinning_reserve}
\end{figure}

\section{Beyond UC-LD problems}

The method proposed here can be easily applied to other problems of different nature.
In order to do this, repair mechanisms need to be modified to deal with the constraints of the given problem.
The modifications, however, are relatively easy and can be done by following the guidelines in Section~\ref{guidelines}.


\section{Conclusions}

In this paper, a method was proposed which can optimize complex electrical dispatch (UC-LD) problems considering all constraints.
Generally speaking, this paper proposes a method able to tackle complex optimization problems with many constraints of varied difficulty.
In fact, it proposes a new constraint handling procedure where parameters of the repair method itself are optimized.
Consequently, this allows (a) the best repair method to be found for the given problem and (b) to have the bias removed.
One of the greatest advantages of the proposed method is that it can be easily applied to other related problems or different settings.
Guidelines were built to aid in this task.





\bibliographystyle{IEEEtran}

\vfill

\end{document}